\documentclass[%
 reprint,
superscriptaddress,
 amsmath,amssymb,
]{revtex4-2}
\usepackage{graphicx}
\usepackage{dcolumn}
\usepackage{bm}
\usepackage{xcolor}
\usepackage{siunitx}

\begin{document}

\title{Metallised 3D printed plastic resonator demonstrates superconductivity below 4 K}

\author{J. Bourhill}
\email{jeremy.bourhill@uwa.edu.au}
\affiliation{Quantum Technologies and Dark Matter Labs, Department of Physics,  University of Western Australia, 35 Stirling Hwy, 6009 Crawley, Western Australia.}
\author{G. Cochet}
\affiliation{Elliptika (GTID) 2 Rue Charles Jourde, 29200 Brest, France}
\author{J. Haumant}
\affiliation{Elliptika (GTID) 2 Rue Charles Jourde, 29200 Brest, France}
\author{V. Vlaminck}
\affiliation{IMT Atlantique, Technopole Brest-Iroise, CS 83818, 29238 Brest Cedex 3, France}
\affiliation{Lab-STICC (UMR 6285), CNRS, Technopole Brest-Iroise, CS 83818, 29238 Brest Cedex 3, France}
\author{A. Manchec}
\affiliation{THALES, 10 Av. 1ère Dfl, 29200 Brest, France}
\author{M. E. Tobar}
\affiliation{Quantum Technologies and Dark Matter Labs, Department of Physics,  University of Western Australia, 35 Stirling Hwy, 6009 Crawley, Western Australia.}
\author{V. Castel}
\affiliation{IMT Atlantique, Technopole Brest-Iroise, CS 83818, 29238 Brest Cedex 3, France}
\affiliation{Lab-STICC (UMR 6285), CNRS, Technopole Brest-Iroise, CS 83818, 29238 Brest Cedex 3, France}
\date{\today}

\begin{abstract}
We report the first observation of a superconducting transition in a 3D printed, metallised-plastic device. A cylindrical cavity is 3D printed from a photosensitive polymer resin and then a 20 $\mu$m layer of tin deposited. A resonant TE microwave mode at 13.41 GHz is observed to reduce its losses by an order of magnitude once it is cooled below 3.72 K; the superconducting transition temperature of tin, with the mode's $Q$ factor increasing from $2.7\times10^4$ to $4.0\times10^5$.

\end{abstract}


\maketitle
\section{Introduction}
The ease, accessibility, cost and versatility of photosensitive liquid resin stereolithography (SLA) 3D printing is vastly superior when compared to traditional subtractive manufacturing techniques. It is not uncommon for 3D printers with micrometer sized print resolutions to be found in the personal home, which is not the case for their cousins; Selective Laser Melting (SLM) 3D printers, which use metallic powders for additive manufacturing, and whose base cost and material cost are substantially higher. Plastic resin 3D printing has also been demonstrated to be an extraordinarily portable technology, with the ability to move a printer from one location to another without the need for arduous set-up and pack-down procedures, or overly cumbersome printing units  {--} a typical printer weighs on the order of 20 kg. This fact makes them an intriguing technology to explore for their applications to defence and space exploration wherein manufacturing from the same device may be required at changing locations with short lead times.

It has been previously demonstrated that microwave resonant cavities with equivalent performance to their traditionally manufactured counterparts can be produced by plastic 3D SLA printing followed by metallisation \cite{8611102,doi:10.1063/5.0006753}. This result has opened up the possibility of rapid, low cost, and highly reproducible production of a wide variety of microwave devices; such as filters, isolators, RF and magnetic field shielding, and more complicated systems for coupling microwave photonics to additional degrees of freedom to form a hybrid system \cite{doi:10.1063/5.0006753,8611102,PhysRevB.107.214423}. 

At the same time, SLM printing has been demonstrated to produce metallic cavities and structures which display superconducting transitions when cooled below their critical temperatures, $T_c$, first in aluminium \cite{doi:10.1063/1.4958684} and then in niobium \cite{9225716}. These advances have been used to construct non-trivial resonant devices \cite{PhysRevD.108.052014,sym14102165} with unique properties which could not have been manufactured through subtractive manufacturing methods, which will greatly benefit from transitioning into superconducting devices at low temperatures. 

Superconducting cavities find numerous uses in physics. Once transitioned, they trap and store resonant microwave radiation and reduce losses, resulting in devices with very high quality factor ($Q$), narrow bandwidth, and long coherence times \cite{doi:10.1063/1.1656986}. These cavities find application in particle accelerators \cite{1440147,PhysRevApplied.13.014024,PhysRevApplied.13.034032}, sensing \cite{PhysRevLett.74.1908,doi:10.1063/5.0023624,Carvalho_2017}, metrology and precision RF sources \cite{PhysRevD.27.1705,STEIN1980363}, and for testing fundamental physics. In particular, tests of the speed of light and the constancy of fundamental constants \cite{PhysRevD.27.1705,PhysRevLett.90.060403,Choi:2021aa,Nagel:2015aa} as well as the search for hidden sector particles and other dark matter candidates \cite{IWAZAKI2020135861,PhysRevD.87.115008,PhysRevD.84.055023}, depend on such cavities. They are also essential in cavity quantum electrodynamics (CQED) experiments to shield qubit devices, thereby providing a reduced density of states for the qubit to radiate into \cite{PhysRevB.86.100506,PhysRevLett.107.240501}. Superconducting cavities are often precision-machined from extremely high purity aluminium or niobium at great material and labour costs in order to achieve optimal surface preparation.

Here, we demonstrate superconductivity in a tin (Sn) coated, 3D-printed-plastic microwave resonant cavity exhibited via the sharp increase of the $Q$ factor of the cavity{'}s lowest-order transverse magnetic mode (TE$_{011}$) at the $T_c$ of Sn, $3.72$ K \cite{DEHAAS1935453}. Sn is a type-I superconductor and has an intermediate $T_c$ located below that of niobium (Nb, $T_c=9.2$ K), but above Aluminium (Al, $T_c=1.2$ K) {--} the two most commonly used elemental superconductors. Given that most commercially available liquid helium based cryogenic systems can reach temperatures below 4 K, often down to or below at least 3.5 K, Sn offers an advantage over Al given it can reach superconductivity in a standard helium{--}4 system without the need for dilution refrigeration, whilst its cost, availability and ease of machinability give it certain advantages over Nb. Sn is readily available and specimens of very high purity can be prepared. If necessary they can be grown as single crystals \cite{RevModPhys.26.277}. These factors make it one of the most convenient superconducting materials.

Whilst the observed $Q$ factor of the resonant TE$_{011}$ mode at mK temperatures is by no means groundbreaking, this study demonstrates an important proof-of-concept of manufacturing superconducting devices from plastic 3D printed structures coated in metal. There are no real limitations other than those imposed by electrochemistry on the type of metal that can be used to coat the plastic or the thickness of this layer. In fact, there exists a great deal of parameter space unexplored in these types of devices in order to optimise the achieved conductivity and hence $Q$ factor. 

\begin{figure}[b!]
\includegraphics[width=0.5\textwidth]{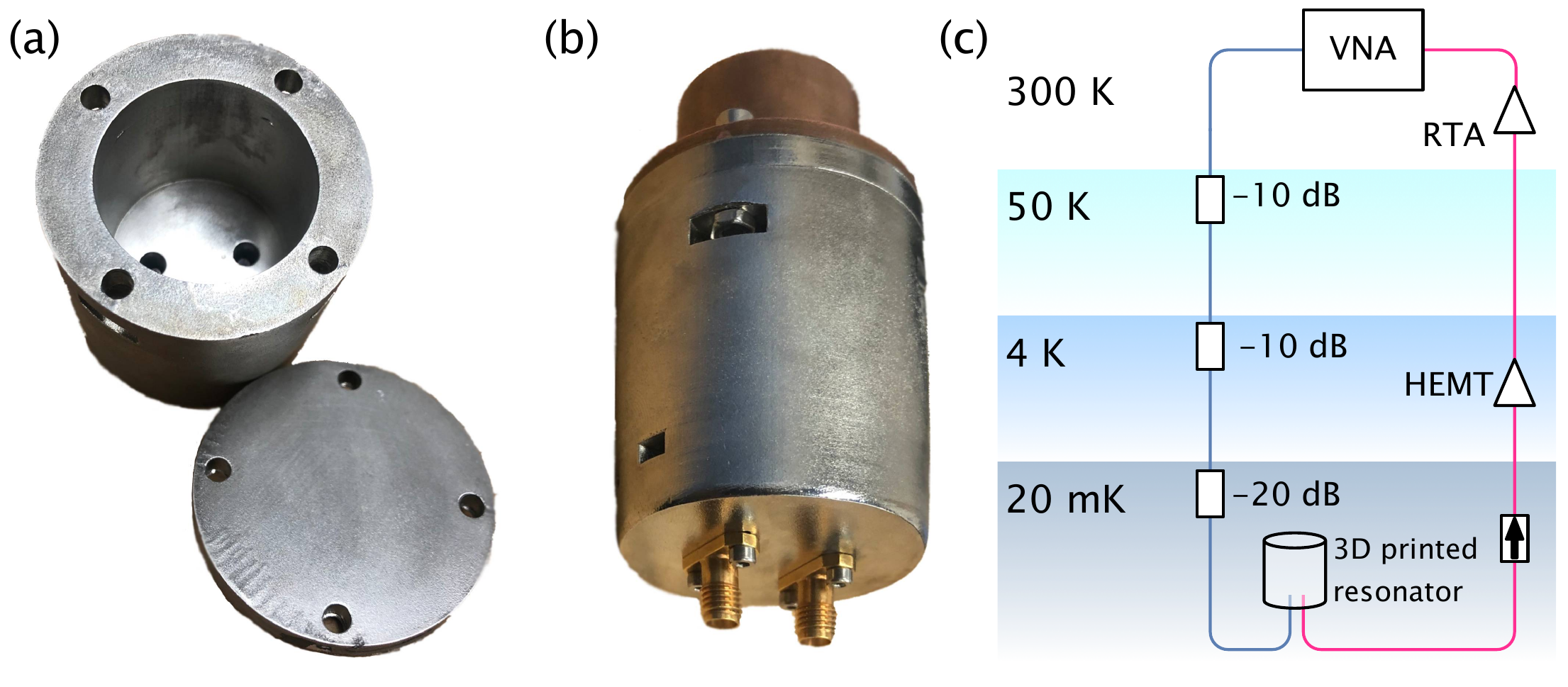}
\caption{The 3D printed, metallised plastic can with internal dimensions $h= 30$ mm, $D=29.375$ mm (a) disassembled as lid and body, and (b) fully assembled with flanged SMA input probes and copper cold-finger attachment. (c) Schematic illustration of the experimental setup in the dilution refrigerator demonstrating cold attenuation on the input line and isolator and amplification (HEMT and room temperature) on the output line.}
\label{fig:res}
\end{figure}

\section{System Description}

\subsection{Cavity manufacturing method}

A hollow cylinder is first 3D printed with the FormLabs$^{\mbox{\scriptsize{\textregistered}}}$ Form 3B+$^{\mbox{\scriptsize{\textregistered}}}$ SLA 3D printer, using the BioMed{--}Clear{--}Resin$^{\mbox{\scriptsize{\textregistered}}}$. This printer has a 25 $\mu$m resolution for the vertical axis. The internal diameter and height of the cavity were both designed to be 30 mm, with radial wall thickness 7 mm. It is printed in two pieces - a main body and a lid (see Fig. \ref{fig:res} (a)), with the lid being 5 mm thick and the base of the main body, which has two holes for the coaxial in{--} and output probes, is 15 mm thick. 

After printing, the cavity was cleaned of residual resin in an isopropyl alcohol bath and then treated in an ultraviolet chamber including a heat treatment at 60$^\circ$ C for 60 minutes.  A conventional 3-D metallisation procedure of plastic elements has been adapted in order to produce a full metallisation of the plastic cavity by Elliptika$^{\mbox{\scriptsize{\textregistered}}}$ and described as follows:\\
\\
\indent i) dry etching by sandblasting (increases surface roughness, increasing pores for Pd$^{2+}$ adsorption);\\
\indent ii) surface activation with Pd$^{2+}$ solution \cite{doi:10.1080/00202967.1959.11869775};\\
\indent iii) autocatalytic bath of copper: Pd particles act as catalyst sites and permit the growth of a homogeneous layer of Cu, which spontaneously reaches 3 $\mu$m thickness;\\
\indent iv) standard electrodeposition process of Cu (10 $\mu$m);\\
\indent v) Sn electrodeposition finish of 20 $\mu$m.\\
\\
It should be noted that thicker layers of Cu and Sn are deposited in steps (iv) and (v) when compared to previous implementations of this procedure \cite{8611102}, which is achieved by simply increasing the run times of the electrodeposition stages. The increase in Cu thickness is to ensure that sufficient heat conduction is achieved in the outer metal layer, given that the bulk plastic body of the system is a poor conductor of heat, it is important if we want the conducting walls of the cavity to reach $T_c$ that the cryostat can efficiently conduct heat away from them. The Sn layer is made thicker to ensure it acts as a bulk superconductor, well above the percolation limit \cite{Beutel_2016}. 

Post metallisation, it is measured that the cavity height is $h=29.7$ mm, whilst the cavity diameter is $D=29.24$. Deviations from the design size are to be expected given the metal layer thickness and the 25 $\mu$m resolution of the 3D printer.

\subsection{Experimental method}
Two coaxial cables terminated in SMA flange mounts act as the input and output RF coupling ports, with the cables inserted through the base of the cavity. It is important when setting the probe couplings at room temperature to ensure that they are very weak. This is because as $Q$ increases, so do the losses from the ports, and we do not wish to over-couple to the mode of interest once transitioned, thereby loading it and indirectly lowering its $Q$ value. In fact, as was done in \cite{doi:10.1063/1.4958684}, we use the opposite probe {``}type{''} in order to minimise coupling losses as much as possible: straight coaxial probes parallel to the cylinder axis are used, which should excite $E_z$ fields despite our mode of interest being a TE mode.

The cavity has a rubidium oxide temperature sensor mounted to its base, ensuring direct measurement of the real cavity surface temperature, and together they are mounted to the mixing chamber (MXC) plate of a dilution refrigerator. The input microwave lines are heavily attenuated, whilst the output signal is amplified with a cryogenic Low Noise Factory$^{\mbox{\scriptsize{\textregistered}}}$ amplifier at the 4 K stage, and a room temperature amplifier, as depicted in Fig. \ref{fig:res}(c). This setup ensures good signal-to-noise ratio of the output spectrum of the cavity when measured with a Vector Network Analyser (VNA). Data is recorded during the condense procedure of the dilution fridge, in which the temperature of the MXC changes from $\sim5$ K to $\sim 20$ mK, an ideal window to observe the superconducting transition of Sn.

\section{Results}
 
\begin{figure}[t!]
\includegraphics[width=0.45\textwidth]{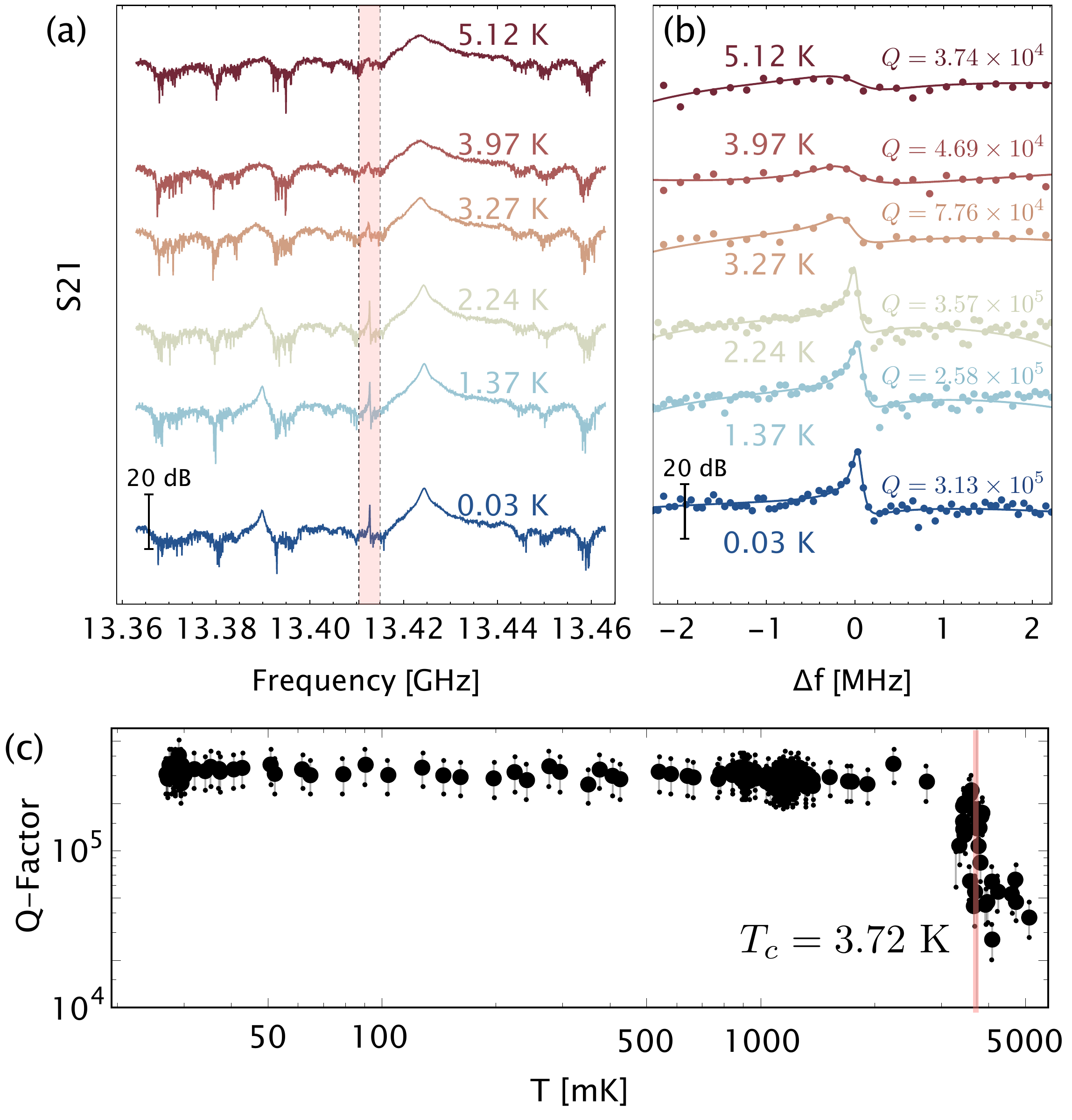}
\caption{(a) S21 transmission spectra over a 100 MHz span at various temperatures with the central highlighted region zoomed in (b). (c) $Q$ factor of the mode extracted from fitting as a function of temperature with the superconducting transition temperature of Sn labelled in red.}
\label{fig:qvt}
\end{figure}

A sample of the $S_{21}$ transmission spectra recorded at different temperatures is displayed in Fig. \ref{fig:qvt}(a) and (b). It can be observed from the first figure that within the observed 100 MHz frequency span there exists 3 modes, each of which becomes more resolved as the temperature decreases; a direct result of the increase in conductivity of the conducting boundaries. 

The mode of interest is the central, high $Q$ TE$_{011}$ mode with frequency $\omega/2\pi=13.413$ GHz. It is determined from finite element simulation that this is indeed the mode we are looking at. A zoomed in picture of the mode is shown in Fig. \ref{fig:qvt}(b) as well as theoretical fits using a Fano model \cite{PhysRev.124.1866} and the resulting calculated $Q$ factor. It is clear that an increase in $Q$ as well as peak transmission is associated with a decrease in temperature.

By plotting the fitted $Q$ values against temperature in Fig. \ref{fig:qvt}(c), we observe that a jump in $Q$ occurs around  $T_c=3.72$ K; the superconducting transition for Sn. Over an order of magnitude improvement in $Q$ factor is achieved; rising from a $T>T_c$ value of $Q\sim2.7\times 10^4$ to a $T<T_c$ value of $Q=4.0\times10^5$.

Being a type-I superconductor, Sn has a typically low critical magnetic field value of 0.03 T \cite{RevModPhys.26.277} above which the superconducting properties of the metal are lost. Indeed, in a second experimental run in which the cavity is placed inside an American Magnetics\textsuperscript{\texttrademark} superconducting magnet mounted to the 4K plate of the fridge, simply switching the magnet on instantly reduces the mode amplitude and $Q$-factor due to some residual magnetisation of the magnet upon power up. \\

\section{Discussion}
A microwave resonance inside an ideal empty metallic structure will be dominated by surface resistance losses. The impact of these losses can be calculated from the so called {``}geometric-factor{''} $G$, where
	\begin{equation}
	G=\mu_0\omega\frac{\iiint|\vec{H}^2|dV}{\iint |\vec{H}_\tau^2|dS},
	\label{eq:gfactor}
	\end{equation}
where $\mu_0$ is the vacuum permeability, $\omega$ is the resonant angular frequency, $\vec{H_\tau}$ is the tangential magnetic field of the resonant mode, $S$ is the surfaces of the resonator, $\vec{H}$ is the magnetic field of the resonant mode and $V$ the cavity volume. The quantity $G$ is related to the $Q$ factor via $G=Q R_s$, where $R_s$ is the surface resistance in ohms. Essentially, the expression $G$ characterises the ratio of the resonant mode{'}s magnetic field within the volume of the cavity compared to that at the surface, where it induces current in the metallic walls and hence experiences resistive loss. 

The geometric factor $G$ obtained through the FEM is approximately 389 $\Omega$ at 13.413 GHz, allowing us to estimate the surface resistance at 972 $\mu\Omega$ below $T_c$ and 14.4 m$\Omega$ above $T_c$. This superconducting resistivity is of equal order to previously achieved values in SLM 3D printed Al and Nb resonators \cite{doi:10.1063/1.4958684,9225716}, however an order of magnitude greater than values measured in 500 nm thick Sn samples \cite{Beutel_2016}.

The measured $Q$ factor is likely limited by two factors; surface roughness and electrical contact between the lid and the cavity. The former could be improved by polishing the Sn layer of the cavity, either mechanically or electrochemically. Conductivity between the two segments of the resonator would likely also be improved with a smoother surface finish of the connecting surfaces. 

In our results we note that on either side of the highlighted mode are two more strongly coupled TM modes {--} their transmission amplitudes are much larger than the central mode, and their $Q$ factors lower as a result of higher $G$ and coupling losses. 

\section{Conclusion}

In conclusion, our study marks a decisive advance in the combination of superconductivity and additive manufacturing. We have successfully demonstrated the first-ever superconducting transition in a 3D-printed metallized plastic device. Cooling the cavity below 3.72 K, the superconducting transition temperature of tin, resulted in a substantial reduction in losses and a noticeable increase in the $Q$ factor, from $2.7\times10^4$ to $4.0\times10^5$, at a TE mode resonant at 13.413 GHz. This breakthrough offers huge potential for any application or industry that requires fast, on-site, short lead-time manufacturing. The adaptability and versatility of this technology offers a promising route to meeting the dynamic demands of certain industries and the access to a  new design parameter space for superconducting devices, unburdened by the limitations of subtractive manufacturing.

\section*{Acknowledgments}
This work was jointly funded by the Région Bretagne through the project OSCAR-SAD18024, by the UWA Research COLLABORATION AWARDS (RCA) grant {``}Investigation of 3D printed microwave cavities at cryogenic temperature{''}, and by the Australian Research Council Centre of Excellence for Engineered Quantum Systems, CE170100009 and the Centre of Excellence for Dark Matter Particle Physics, CE200100008..

\bibliography{can_bib.bib}

\end{document}